\def\P{$P$}

\def\eq{\enskip =\enskip}
\def\pls{\enskip +\enskip}
\def\mns{\enskip -\enskip}

  \def\ket{\vert \vert  \{ \emptyset \} \rangle}
  \def\ket2{\vert \vert \otimes \{ R \} \rangle}

\def\.#1{\mathaccent 95#1}
\def\^#1{\mathaccent 94 #1}
\def\~#1{\mathaccent "7E #1}

\def\eq{\enskip =\enskip}
\def\pls{\enskip +\enskip}
\def\mns{\enskip -\enskip}

  \def\ket{\vert \vert  \{ \emptyset \} \rangle}
  \def\ket2{\vert \vert \otimes \{ R \} \rangle}

\def\be{\begin{equation}}
\def\ee{\end{equation}}

\documentclass[12pt,epsfig]{iopart}
\usepackage{epsfig}
\begin{document}
\title{Magnetic transition in Ni-Pt alloy Systems : Experiment and Theory}
\author{Uday Kumar\dag, K. G. Padmalekha*,   
P. K. Mukhopadhyay\dag\footnote{email:pkm@bose.res.in}
Durga Paudyal**\footnote{email:dpaudyal@bose.res.in} and Abhijit Mookerjee**}
\address {\dag Laboratory for Condensed Matter Physics, 
 S.N. Bose National Centre for Basic Sciences,
JD Block, Sector 3, Salt Lake City, Kolkata 700098, India}
\address { *Department of Physics, Indian Institute of Science, Banglore 560012, India}
\address {** Condensed Matter Theory group, 
 S.N. Bose National Centre for Basic Sciences,
JD Block, Sector 3, Salt Lake City, Kolkata 700098, India}

\date{\today}

\begin{abstract}
We report here the preparation and measurements on the susceptibility, sound velocity and 
internal friction  for Ni-Pt systems. We then compare these experimental results with the 
first principle theoretical predictions and show that  there is reasonable agreement with 
experiment and theory.
\end{abstract}


\section{Introduction}

The Ni-Pt alloy system is an interesting study both because of earlier controversies about 
theoretical predictions regarding its  chemical stability and its
magnetic properties.

Some initial  studies predicted these systems to be phase separating  contradicting 
 experimental observations. 
In our previous work \cite{paud,durga2} we  concluded that NiPt system is stable 
{\sl provided} we take into account scalar relativistic corrections to the underlying
Schr\"odinger equation and deal with both
 charge transfer and lattice relaxation effects properly.

For the 50$\%$ alloy there is also a disagreement regarding its magnetic properties. 
Early experiments indicated that disordered NiPt is ferromagnetic 
\cite{exp1}, while in the ordered phase it is paramagnetic \cite{exp3}. Other authors found ordered NiPt to be
antiferromagnetic \cite{expt2}. Spin polarized local density approximation based calculations
seem to indicate that even Ni$_{25}$Pt$_{75}$ shows some local magnetic moment, whereas
experiments seem to indicate that there is no magnetism at all.

In this communication we report susceptibility, sound velocity and internal friction
experiments on a series of NiPt alloys with Pt concentrations varying between 41$\%$
and 76$\%$. We have also carried out spin-polarized local-density based tight-binding linearized
muffin-tin orbitals (TB-LMTO) calculations for the magnetic properties of these alloys and
analyze the experimental results in this light.

\section{Experimental details}
\subsection{Sample preparation and characterization}

We made four different compositions of the alloy, Ni$_{\mathrm{x}}$Pt$_{\mathrm{1-x}}$ (in atomic percentages). First an amount of Ni was cut from an ingot of pure Ni. After weighing it, a target amount of Pt was cut from pure Pt wire. Since there would always be a little bit of error in weight adjustments, the exact target composition was never reached, but we determined the final composition to be close to it. This is tabulated in table \ref{t1}.

\begin{table}[h]
\caption{Composition analysis for the NiPt samples.}
\begin{center}
\begin{tabular}{cccc}
\br
Target             & Actual             & Nearest whole number   & Lattice Constants \\
composition        & composition        & composition            & in nm \\ \mr
Ni$_{60}$Pt$_{40}$ & Ni$_{58.7}$Pt$_{41.3}$ & Ni$_{59}$Pt$_{41}$ & 0.372 \\
Ni$_{50}$Pt$_{50}$ & Ni$_{49.6}$Pt$_{50.4}$ & Ni$_{50}$Pt$_{50}$ & 0.376  \\
Ni$_{45}$Pt$_{55}$ & Ni$_{44.6}$Pt$_{55.4}$ & Ni$_{45}$Pt$_{55}$ & 0.377 \\ 
Ni$_{25}$Pt$_{75}$ & Ni$_{23.9}$Pt$_{76.1}$ & Ni$_{24}$Pt$_{76}$ & 0.384 \\ 
\mr
\end{tabular}
\end{center}
\label{t1}
\end{table}

	The materials for the required composition were then put in an arc furnace and melted in a flowing argon atmosphere. After melting, the mass formed into a shinning button. We measured the maximum mass loss to be 0.8\%. These are now taken out and initial homogenizations were done under sealed and evacuated quartz ampoules at 1000$^o$C for 12 hours. Then they were quenched to room temperature. Afterwards they were carefully cold rolled to about 0.5mm thickness and cut into reed shapes.

The samples are then again put in evacuated quartz ampules and heat treated to 1000$^o$ C for 72 hours. After quenching, they were finally annealed at 200$^o$ C for 4 hours to remove the stresses due to thermal shock that develops due to fast quenching.

\begin{figure}[p]
\centering
\epsfxsize=5.0in\epsfysize=5.0 in
\rotatebox{0}{\epsfbox{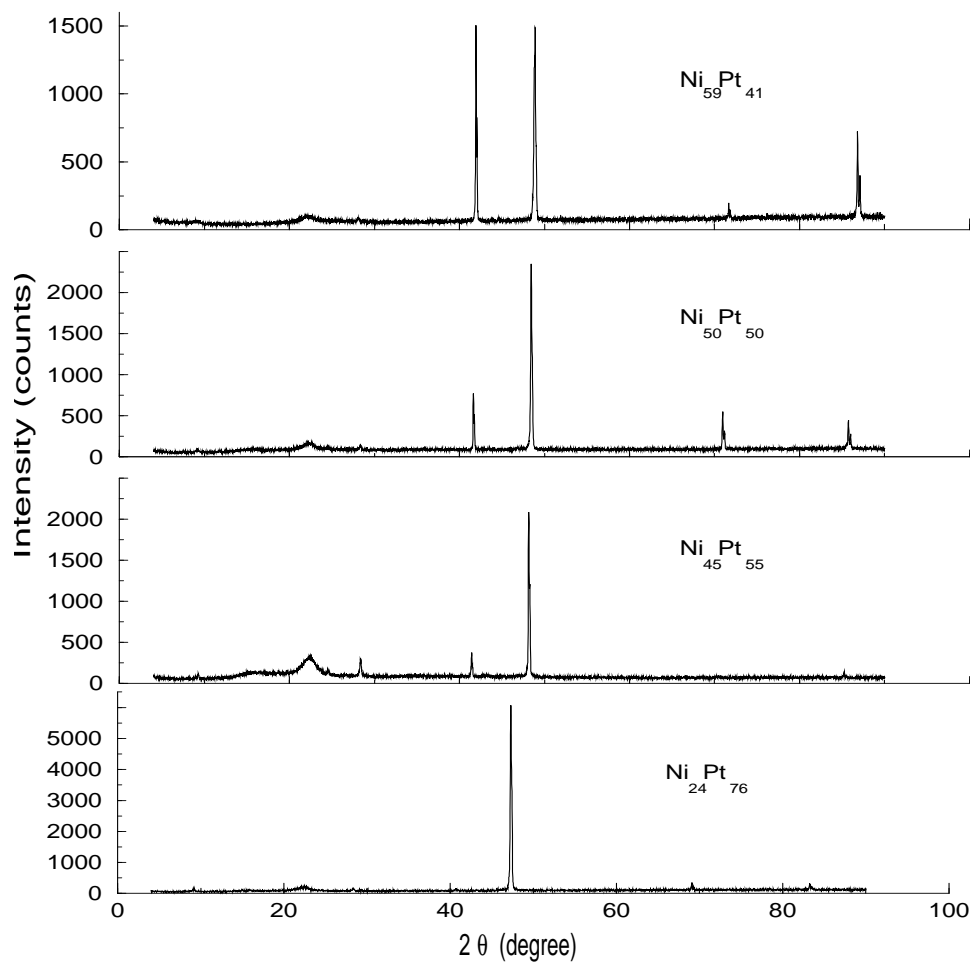}}
\caption{XRD for NiPt systems}
\label{f1}
\end{figure}

The crystal structures of the samples were then measured in a standard XRD instrument (Philips make). Scans were taken from 4$^o$ to 90$^o$ at an interval of 0.02$^o$ with a step time of 0.5 sec. Diffractograms for the samples are shown in figure \ref{f1}.

The analysis of the XRD data indicate that all the samples have signatures of 
 the face centered cubic (fcc)  structure. If the alloy was ordered, it would have showed
the signatures of the L10, L12 or other relevant superstructures. We do not see any indication of that.
This  implies that all the samples 
were probably in the  disordered phase which is expected 
because of  homogenization at and quenching from  1000$^o$ C. 
There is good agreement between our experimental lattice parameters and that of Parra \etal \cite{exp1}. 

\begin{figure}
\centering
\epsfxsize=4.0in\epsfysize=2.5 in
\rotatebox{0}{\epsfbox{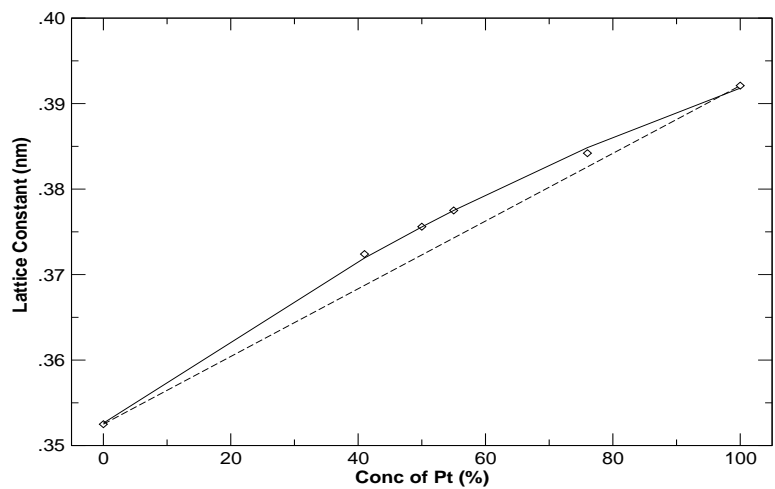}}
\caption{Deviations from Vegard's Law (dashed lines)}
\label{veg}\end{figure}

The figure \ref{veg} shows the lattice constants as a function of the Pt concentration. The figure
clearly shows deviations from  Vegard's Law and a positive `bowing' effect. This is to be
expected because of the large size mismatch between Ni and Pt atoms. The maximum deviation occurs
at around the 50$\%$ concentration.

\subsection{Susceptibility measurements}

\begin{figure}
\centering
\psfig{file=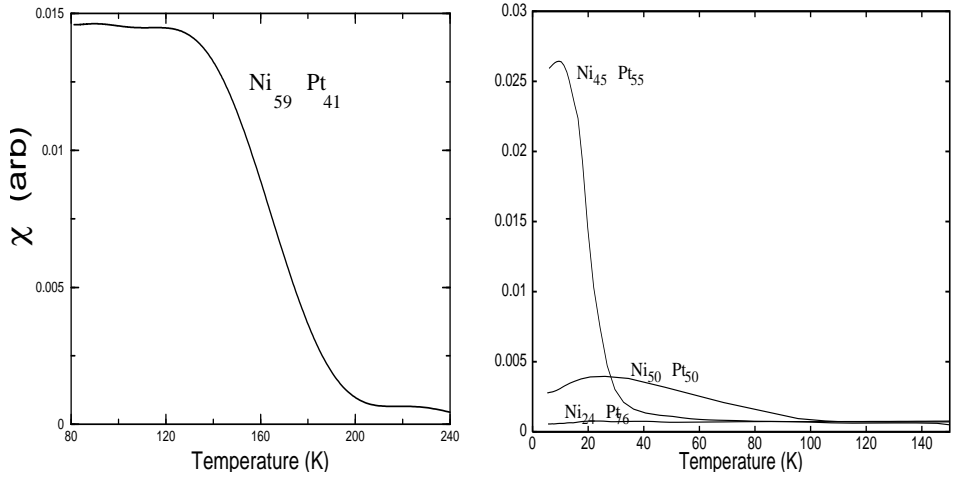,height=2.5in,width=5in,angle=0}
\caption{Susceptibility curve for NiPt with (left) 41$\%$ Pt and (right) 50$\%$, 55$\%$ and 76$\%$ Pt}
\label{sus}
\end{figure}

In figure \ref{sus} we show the data for the susceptibilities of the four samples. All except

\begin{figure}[h]
\centering
\vskip 0.5cm
\psfig{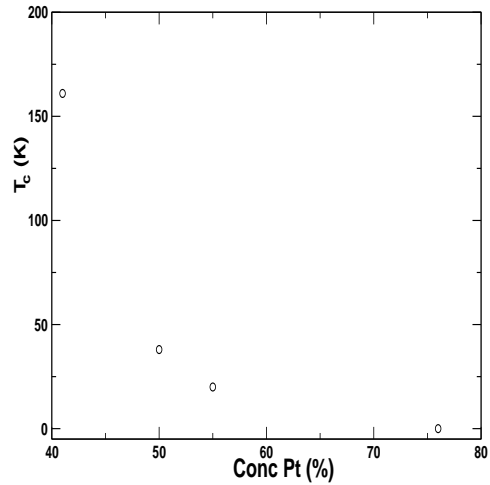}
\caption{Experimental Curie temperature $vs$ the alloy composition}
\label{tc}
\end{figure}

\noindent for Ni$_{59}$Pt$_{41}$ were done till liquid helium temperature. The measurements were done in a standard
double balanced coil technique. Driving field was about 100 Oe and frequency was 120 Hz for the
three samples, whereas for Ni$_{59}$Pt$_{41}$ it was 33 Hz. Taking T$_c$ to be the temperature where $\partial \chi$/$\partial$T 
shows a maximum, we find that for Ni$_{59}$Pt$_{41}$ T$_c$ is  around 161K. We summarize the various $T_c$s in figure \ref{tc}. 

\subsection{Sound velocity and attenuation}

\begin{figure}
\centering
\psfig{file=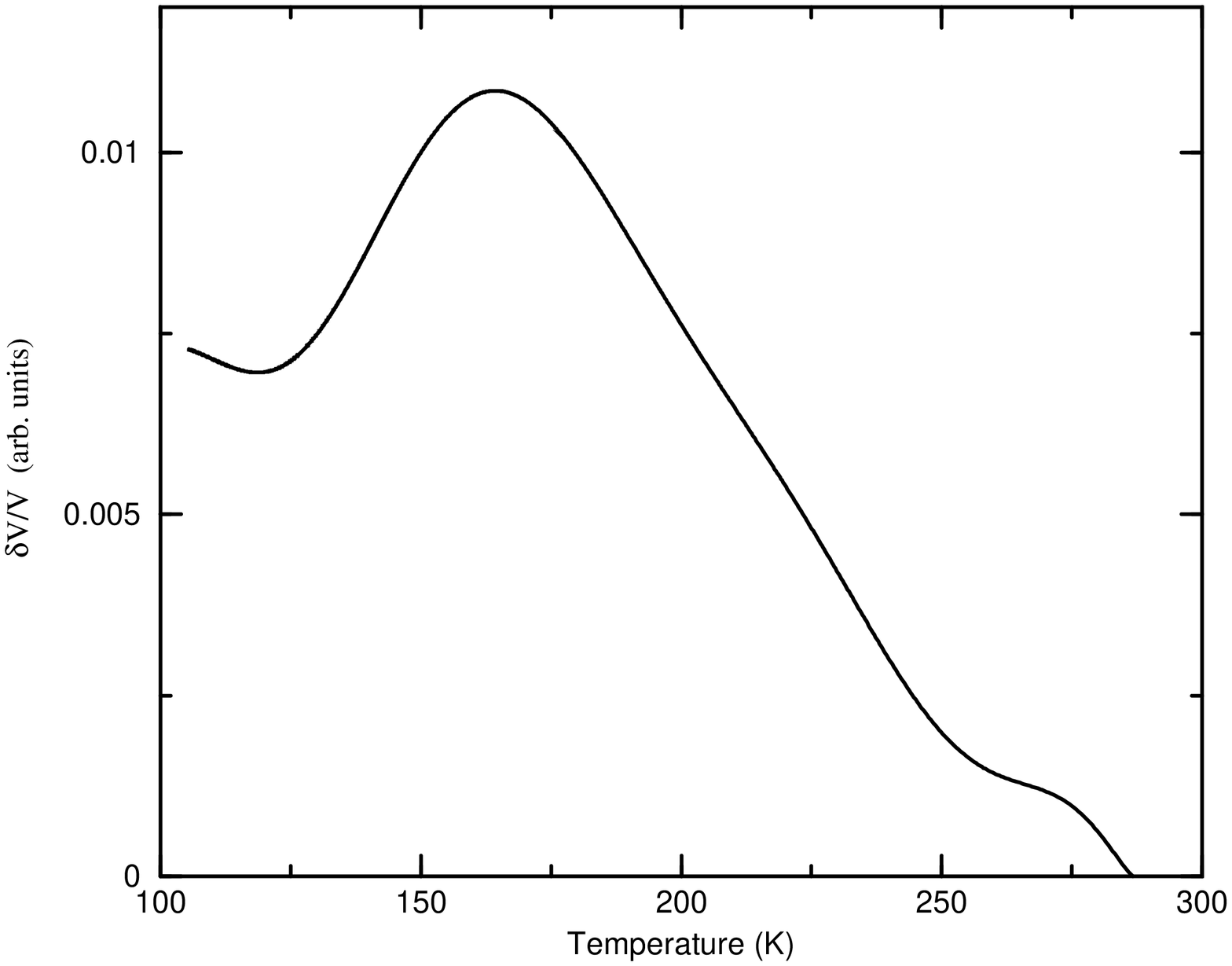,height=10cm,angle=270}
\psfig{file=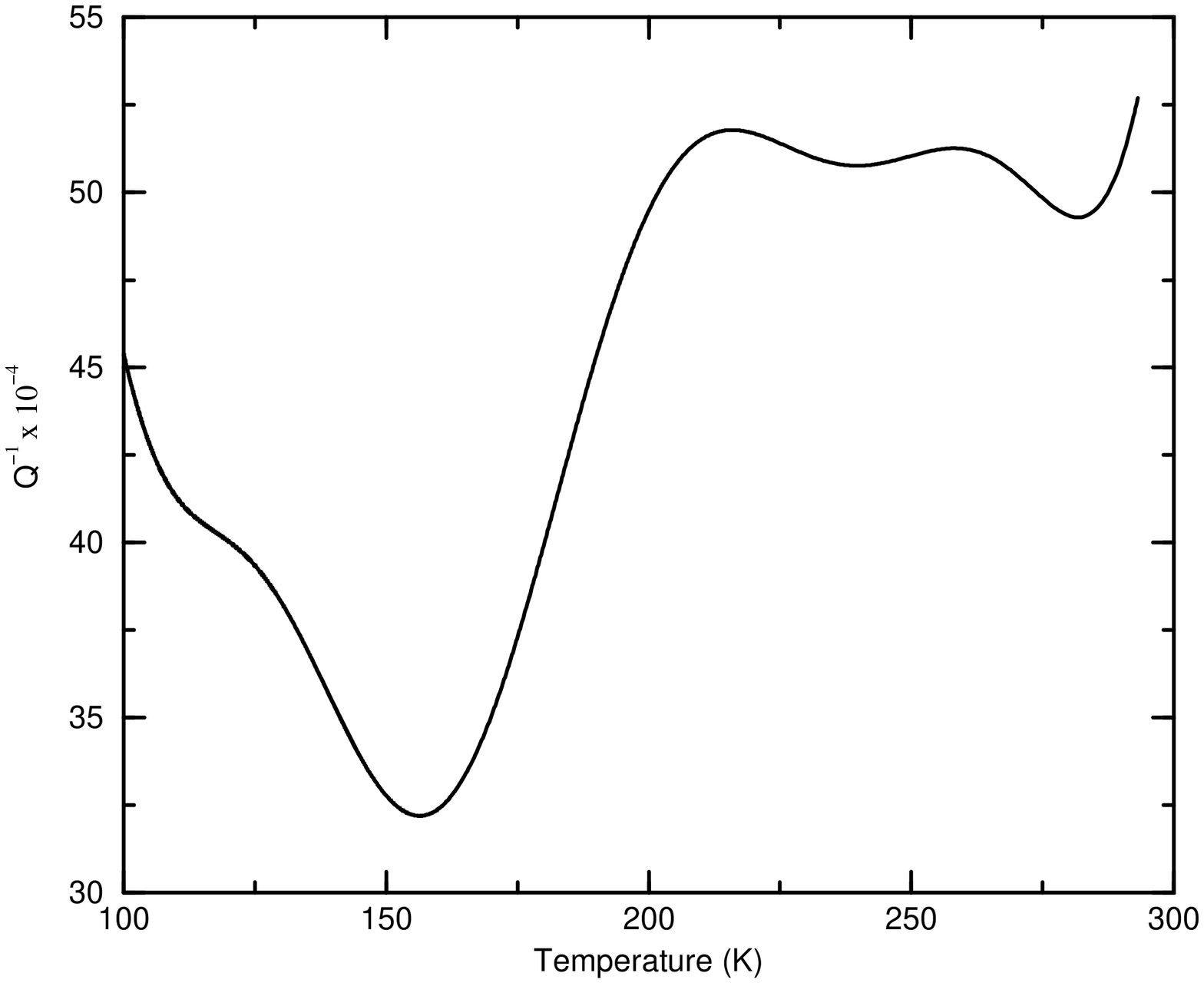,height=10cm,angle=270}
\caption{(Top) Sound velocity  and  
 (Bottom) Internal friction for Ni$_{59}$Pt$_{41}$}
\label{f4}
\end{figure}

In this section we describe the sound velocity and internal friction (Q$^{-1}$) measurements done on the systems as a function of temperature. However, since the present setup does not reach below about 77K, data for only the Ni$_{\mathrm{59}}$Pt$_{\mathrm{41}}$ are shown in figure 5. The top figure shows the sound velocity results and the bottom one shows the internal friction data. In this case we could measure to temperature lower that T$_{\mathrm{c}}$.

These measurements are done by the vibrating reed technique. For this the sample was taken in the form of a reed and clamped at one end in the sample holder. The free end was made to resonate by electrostatic drive from one driver electrode placed near one surface there. Another electrode at the opposite side picked up this signal and fed to a lock-in amplifier (LIA). This LIA was run at 2f mode from the reference provided by the signal generator for the driver. An integrator completed the phase locked loop that was setup to track the changes in resonance characteristics of the sample as the temperature was varied. The temperature was measured with the help of a platinum resistance thermometer. Detailed description of the vibrating reed technique can be found in \cite{kn:vi_reed1}.

From figure 5 (top), we see that as the temperature was lowered, initially the sound velocity was increasing. However, as the temperature approached T$_{\mathrm{c}}$, the lattice softened up and the sound velocity showed a good amount of drop. However, for T far below T$_{\mathrm{c}}$, the sound velocity again picked up, signaling the end of the magnetic transition and returned to the background lattice elastic limit. In case of internal friction (figure 5, bottom), we see an exactly complementary behaviour. Here the general lowering of Q$^{\mathrm{-1}}$ was enhanced as T approached T$_{\mathrm{c}}$, till it dropped to about 50\% of the room temperature value. And just like the sound velocity, it tended to recover to a general lower value at still lower temperatures.

Both these elastic measurements show that with the onset of magnetic transition, the background lattice elastic properties get largely  modified due to strong spin lattice coupling in this system. In case of sound velocity which is the real part of elasticity, the elastic modulus fell to  about 50\% across the transition. The peak temperature, T$_{\mathrm{p}}$, is about 161K which is about the same as the T$_{\mathrm{c}}$ as determined from the real part of $\chi$. However, it is well known that the effect of ferromagnetism on spin lattice coupling can start from a little below 2T$_{\mathrm{c}}$ \cite{kn:vi_reed2,kn:vi_reed3}, we actually see that the sound velocity started a faster rise below around 260K. The same argument predicts reduction of effect of magnetic transition down from T$_{\mathrm{c}}$/2 and it is also seen in the recovery of sound velocity.

In the similar vein, we can see the effect of magnetic transition on Q$^{\mathrm{-1}}$ data too. We see that the general fall of Q$^{\mathrm{-1}}$ was arrested below 250K and a massive fall in it occurred. The minimum temperature, T$_{\mathrm{m}}$, is about 158K -- below T$_{\mathrm{p}}$. Again this is well documented \cite{kn:vi_reed2} that T$_{\mathrm{m }}$(determined from the imaginary part) always lags behind T$_{\mathrm{p }}$(determined from the real part). And just like the real part, Q$^{\mathrm{-1}}$ also recovered from the effects of magnetic transition below about T$_{\mathrm{c}}$/2. Hence we find that the effect of magnetic transition was to both soften the lattice and also reduce the anharmonicity of it simultaneously. We could have tried to fit a general elastic background to the data and extract the magnetic part from it. However, there are major problems in it. We need to have data from much above 2T$_{\mathrm{c}}$ to much below T$_{\mathrm{c}}$/2 -- which are not available. And in any case the fittings are always ambiguous and subject to intense debate. So we would not like to do it here.

\section{Theoretical and Computational methods}

For a theoretical estimation of the Curie temperatures we have followed two different approaches.
 In the first method, we  calculate the 
magnetic pair energies and then use the mean field approximation to evaluate the approximate Curie temperatures.
For magnetic pair interaction energies we  embed different atoms with their accompanying spins
in an  averaged non-magnetic disordered medium  at two different sites a distance
R apart. The pair interactions
are defined as :

\begin{eqnarray*}
J_{AA}(R) = E^{\uparrow \uparrow}_{AA} + E^{\downarrow \downarrow}_{AA} - E^{\uparrow \downarrow}_{AA}
- E^{\downarrow \uparrow}_{AA}\nonumber\\
J_{BB}(R) = E^{\uparrow \uparrow}_{BB} + E^{\downarrow \downarrow}_{BB} - E^{\uparrow \downarrow}_{BB}
- E^{\downarrow \uparrow}_{BB}\nonumber\\
J_{AB}(R) = E^{\uparrow \uparrow}_{AB} + E^{\downarrow \downarrow}_{AB} - E^{\uparrow \downarrow}_{AB}
- E^{\downarrow \uparrow}_{BB}
\end{eqnarray*}

We obtain these pair interactions directly using the orbital peeling technique \cite{op}
in conjunction with the augmented space recursion \cite{asr} introduced by us earlier
in a tight-binding linearized muffin-tin orbitals basis (TB-LMTO) \cite{lmto}.

Using these interaction energies the Bragg-Williams mean-field expression for Curie temperature is 

\[T_c = \frac{1}{k_B}\ \left[ \rule{0mm}{4mm}- x^2 J_{AA} - (1-x)^2 J_{BB} + 2x(1-x) J_{AB}\right]\]

\noindent $J_{AA} = \sum_R\ J_{AA}(R)$ etc. 

The other approach is the Mohn-Wohlfarth (MW) procedure \cite{mw} :

$$  \left(\frac{T_{C}}{T_{S}}\right)^2+\frac{T_{C}}{T_{SF}} \mns 1 \eq 0$$
where,
T$_S$ is the Stoner  temperature calculated from the relation
$$
\langle I(E_F)\rangle  \int_{-\infty}^{\infty}\ dE\ N(E)\ \left(\frac{\partial f}{\partial E}\right) 
\ =\ 1
$$
$\langle $I(E$_{F}$)$\rangle$ is the concentration averaged Stoner parameter 
, $N(E)$ is the density of states per atom per spin of the paramagnetic state \cite{gun}
and $f(E)$ is the Fermi distribution function.
The spin fluctuation temperature $T_{SF}$ is given by,
\[
T_{SF} \eq \frac{m^2}{10k_{B} \langle \chi_{0}\rangle}
\]
$ \langle \chi_{0} \rangle$ is the concentration weighted exchange enhanced spin susceptibility at 
equilibrium and $m$ is the averaged magnetic moment per atom.
$\chi_{0}$ is calculated using the relation by Mohn \cite{mw} and Gersdorf
\cite{ger}:

\[\chi_{0}^{-1} \eq  \frac{1}{2\mu_{B}^2}\left(\frac{1}{2N^\uparrow(E_{F})}\pls
\frac{1}{2N^\downarrow(E_{F})} \mns I\right)\]

\noindent $N^\uparrow(E_{F})$ and $N^\downarrow(E_{F})$ are
the spin-up and spin-down partial density of states per atom at the Fermi level for each species in the alloy.

\subsection{Lattice Structure}

\begin{table}
\centering
\begin{tabular}{ll}
\br
Alloy & Lattice \\
Composition & Constant (nm) \\ \mr
Ni$_{60}$Pt$_{40}$ & 0.372 \\
Ni$_{50}$Pt$_{50}$ & 0.376 \\
Ni$_{45}$Pt$_{55}$ & 0.378 \\
Ni$_{25}$Pt$_{75}$ & 0.384 \\
\br
\end{tabular}
\caption{Equilibrium lattice constants in nm} 
\label{t2}
\end{table}

As shown in the table \ref{t2} and in comparison with table \ref{t1}, our measured values of lattice parameters match well with the previously 
measured lattice parameters. In all concentrations we have got the FCC structures as par with the 
previous experiment by Parra \etal \cite{exp1}. It should be noted that our calculations show that
for the ordered 50$\%$ alloy, the structure with a small tetragonal distortion has the lowest energy.
 The experimental lack of signature of any tetragonal distortion
for the 50$\%$ alloy further strengthens our belief that the alloy was indeed disordered.
 
\subsection{Curie Temperature}

The experimental measurements in NiPt alloys with 40$\%$, 50$\%$ and 55$\%$ concentration of Pt show the 
ferromagnetic to paramagnetic transitions are  at 161, 38 and 20K.

Experimentally we find that the alloys up to 55$\%$ Pt do show ferromagnetic behaviour, certainly
in the disordered phase. However, at 76$\%$ of Pt the alloy becomes non-magnetic. Our first
principles calculations agree with this up to 55$\%$ of Pt. However, our theoreical 
calculations indicate that ferromagnetism persists even at 75$\%$ of Pt. We are unable to
explain why this should be and why the Stoner-like arguments fail in these compositions.
Table \ref{t3} shows that the Mohn-Wohlfarth and Bragg-Williams approaches both 
 give qualitatively the correct trend in T$_c$ up to 55$\%$ of Pt.
Both approaches show the 50$\%$ alloy to be ferromagnetic, settling an earlier dispute.
It is interesting to note that if we do not take into account local lattice distortion due
to size mismatch in the NiPt alloys, the 50$\%$ alloy yields a Curie temperature of 76K 
which is quantitatively at variance with experiment. We have included size mismatch effects
within our TB-LMTO-Augmented Space Recursion.

\begin{table}[h]
\caption{Curie temperatures in Ni-Pt alloys.}
\begin{center}
\begin{tabular}{cccc}
\br
Composition             & Experiment & Theory I     & Theory II     \\
                        &            & (Mohn-Wohlfarth)  & (Bragg-Williams)   \\ \mr
Ni$_{60}$Pt$_{40}$      & 161        & 125                & 232                 \\
Ni$_{50}$Pt$_{50}$      &  38        & 38                 &  40                 \\
Ni$_{45}$Pt$_{55}$      &  20        & 35                 &  17                 \\
\br
\end{tabular}
\end{center}
\label{t3}
\end{table}

\section{Conclusions}
 In this communication we have reported structural and magnetic properties of NiPt alloys. The structure
was probed by XRD experiments, while the magnetic properties were probed by susceptibility, sound velocity
and attenuation. Our experimental findings of lattice parameters agree with earlier experiments
and confirm that our samples were in the disordered state. We have also carried out a theoretical study of the same alloy system and compared the
theoretical predictions with the experimental results. The theory correctly predicts the trend
in the Curie temperature. The $Ni_{50}Pt_{50}$ alloy was found to be ferromagnetic, settling an
earlier dispute. Our theory cannot explain why ferromagnetism disappears around the $Ni_{25}Pt_{75}$ composition.

\ack
\vskip 0.3 cm
\noindent We would like to thank Drs. A. Mitra and P. K. De for the samples, Prof.A.K.Majumder for the use of his
arc furnace and Prof. A.K. Raychaudhuri for measurement till 5K. We would also like to thank Dr. D. Das
for his various helps at critical stages and Mr. S. Chanda for help in rolling the samples. We would like to thank Mr. D. Hazra for his help with XRD analysis. We would like to
acknowledge partial support from TWAS under its project grant 97-212 RG/PHYS/AS and the CSIR project grant 
03(948)/02-EMR-II.

\end{document}